\documentclass[11pt]{article}
\usepackage{moriond,epsfig}

\def\be{\begin{equation}}
\def\ee{\end{equation}}
\def\bea{\begin{eqnarray}}
\def\eea{\end{eqnarray}}

\begin{document}
\vspace*{4cm}
\title{ Planned Dark Matter searches with the MAGIC Telescope }

\author{ JOSEP FLIX for the MAGIC Collaboration}

\address{ Institut de F\'{\i}sica d'Altes Energies \\
Edifici Cn, Universitat Autonoma de Barcelona, 08193, Spain
}

\maketitle\abstracts{The MAGIC 17m-diameter Imaging Air Cherenkov Telescope
(IACT) has been commissioned beginning of 2005. The telescope has been
designed to achieve the lower detection energy threshold ever obtained with an
IACT, about 50 GeV. A new window in $\gamma$-ray astronomy is being opened with
great impact for exciting new physics and new discoveries. Among the targets
of MAGIC is the indirect detection of Dark Matter (DM). We have
considered different DM halo models of high DM density objects like the
center of the Milky Way, its closest satellites and nearby galaxies
(M31,M87). For each object, detection limits are computed for different DM
halo models in a mSUGRA scenario for supersymmetric neutralino
annihilation $\gamma$-ray production. Advantages and drawbacks of these
objects and plans for future observations are discussed.}

\section{Introduction}

At present, the nature of Dark Matter is still unknown, and none of the known
particles satisfy all requirements to account for it. A number of viable
Weakly Interacting Massive Particle (WIMP) candidates have been proposed
within several theoretical frameworks, mainly motivated by extensions of the
standard model of particle physics, e.g. supersymmetry (SUSY). Among this
variety of particles, the most plausible candidate is the neutralino~\cite{jung96}
($\chi$).

Any WIMP candidate (SUSY or not) may be detected directly via elastic
scattering with targets in the Earth. There are several dedicated
experiments already exploiting this detection technique, but they
do not claim any strong and solid detection up to now (see
review~\cite{gas05}). Complementary, neutralinos might be indirectly detected
by their self-annihilation products in high-density Dark Matter
environments. In particular, channels that produce gamma-rays are interesting
because $\gamma$-rays are not deflected by magnetic fields and preserve the
information of the original annihilation region, i.e. they act as tracers of
the Dark Matter density distribution. This continuum gamma-ray spectra may be
indirectly observed by means of IACTs.

\section{The MAGIC Experiment}

To date, the Major Atmospheric Imaging Cerenkov telescope (MAGIC~\cite{cor05}) is the largest
world-wide Imaging Air Cerenkov Telescope (IACT). Located on the Canary Island La Palma
(2200m a.s.l), the telescope has a 17m diameter high reflectivity tessellated
parabolic mirror dish, mounted on a light weight carbon fiber frame. It is
equipped with a high efficiency 576-pixel photomultiplier camera, whose
analogue signals are transported via optical fibers to the trigger electronics
and the 300 MHz FADC readout. Its physics program comprises, among other topics,
pulsars, supernova remnants, active galactic nuclei, micro-quasars, gamma-ray
bursts and Dark Matter. 

Several positive detections of already known $\gamma$-ray emitters
have been already accomplished during its initial phase (see contributions to
these proceedings from N. Tonello [1ES1959+650], M. L\'{o}pez [Crab Nebula]
and D. Mazin [Mrk-421]). Coping with the hadronic background below 100 GeV
presents a new challenge, but these observations evidence that MAGIC analysis
extends well below 100 GeV and, encouragingly enough, even $\gamma$-rays with
an estimated energy of 50 GeV trigger the telescope. The analysis methods
are presently being adapted to the low-energy domain, uncharted territory as
yet. Meanwhile, the collaboration is involved in the construction of a second
telescope (MAGIC-II, see A. Moralejo contribution to these proceedings). This
will improve the angular and spectral resolution and flux sensitivity of the
system.

\section{Gamma-rays from neutralino annihilations}

Neutralino annihilation can generate continuum $\gamma$-ray emission,
via the process $\chi\chi\rightarrow q \bar{q}$. The subsequent decay
of $\pi^{0}$-mesons created in the resulting quark jets produces a continuum
of $\gamma$-rays. The expected annihilation $\gamma$-ray flux above an energy
$E_{thresh}$ arriving at Earth is given by:

\begin{equation}
\frac{dN_{\gamma}(E_{\gamma}>E_{\mathrm{thresh}})}{dt\ dA\ d\Omega } = N_{\gamma}(E_{\gamma}>E_{\mathrm{thresh}}) \cdot \frac{1}{2} \cdot \frac{\langle \sigma v \rangle}{4 \pi m_{\chi}^2}\cdot \int_{los}\rho_{\chi}^2(\vec{r}(s,\Omega)) ds\ ,
\end{equation}

where $\langle \sigma v \rangle$ is the thermally averaged annihilation cross
section, $m_{\chi}$ the mass and $\rho_{\chi}$ the spatial density
distribution of the hypothetical Dark Matter
particles. $N_{\gamma}(E_{\gamma}>E_{\mathrm{thresh}})$ is the gamma yield
above the threshold energy per annihilation. The predicted flux depends on the
SUSY parameters and on the spatial distribution of the Dark Matter. The energy
spectrum of the produced gamma radiation has a very characteristic feature
with a sharp cut-off at the mass of the Dark Matter particle. Moreover, the
flux should be absolutely stable in time.

High density Dark Matter regions are the most suitable places for indirect
Dark Matter searches. From simulations and stellar dynamics we know that
they correspond mainly to the central part of galaxies and Dark
Matter dominated dwarf-spheroidal-satellite galaxies (with large mass-to-light
ratio). Numerical simulations in a Cold Dark Matter framework predict a few
universal DM halo profiles (for example see~\cite{nfw97}). All of them differ
mainly at low radii (pc scale), where simulation resolutions are at the very
limit. 

Combining the SUSY predictions with the models of the DM density profile
for an specific object, the gamma flux from neutralino annihilations can be
derived. Concerning SUSY, we made use of a detailed scan made in the Minimal
Supergravity (mSUGRA), a simple and widely studied scenario for supersymmetry
breaking. Details of the scan are explained in Prada et al~\cite{Prada04}. For
a given choice of SUSY parameters, $m_{\chi},\;\langle \sigma v \rangle$ and
$N_{\gamma}$ are determined, consistent to all observational constrains. High
DM density objects like the center of the Milky Way, its closest satellites
and the nearby galaxies (M31,M87) are prime candidates for the
observational study.

\subsection{Galactic Center}

The presence of a Dark Matter halo in the Milky Way Galaxy is well established by
stellar dynamics. In particular, stellar rotation curve data of the Milky Way
can be adjusted by fitting with the use of universal DM profiles predicted by
simulations. The most recent models include an effect that had been previously
neglected and affects the DM density at the center of the Milky Way halo,
namely the compression of the Dark Matter due to the infall of baryons to the
innermost region~\cite{Prada04}. As Dark Matter density is enhanced at the
center the expected signal is boosted. This model is based on a large amount
of observational data of our galaxy and is in good agreement with the
brand new cosmological simulations for the response of Dark Matter halos to
condensation of baryons~\cite{gnedin04}.

Such a central spike in the center of the galaxy is under debate
and depend on the Black Hole history during galaxy formation. There exist some
evidences against a central cusp, but they must be taken with caution (see
F. Ferrer in these proceedings). We consider an uncompressed NFW DM halo
model~\cite{fornengo04} and the adiabatic contracted NFW profile~\cite{Prada04}.

\subsection{Draco dwarf spheroidal and nearby galaxies}

The Milky Way is surrounded by a number of small and faint companion
galaxies. These dwarf satellites are by far the most Dark Matter dominated known
objects, with Mass-to-Light ratios from 30 to 300. Draco is the most DM
dominated dwarf satellite. DM density profiles derived from Draco stars cannot
differentiate between cusped or cored profiles in the innermost region, as
data are not available at small radial distances. Moreover, observational data
disfavors tidal disruption effects, which may affect dramatically the DM
distribution in Draco. In order to compare with the Galactic Center we adopt
the very recent cusped DM model which includes new available Draco
data~\cite{Lokas05}. 

In addition, we adopted NFW models for the nearby galaxy
M31~\cite{evans04} and the Virgo Cluster~\cite{mclaugh99}. These NFW profiles do not
take into account any enhancement effect, like adiabatic contraction or
presence of DM substructures.

\begin{figure}
\begin{center}
\psfig{figure=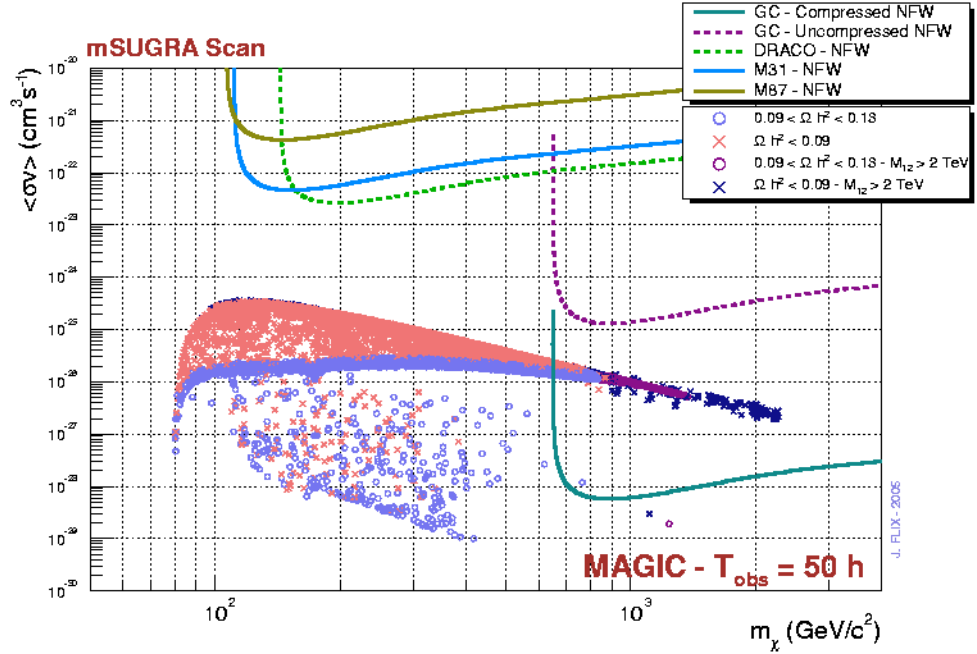,height=3.5in}
\caption{Exclusion limits for the four most promising sources of Dark Matter
annihilation radiation. The Galactic Center is expected to give the largest flux (lowest
exclusion limits) amongst all sources.
\label{fig:exclusion_limits}}
\end{center}
\end{figure}

\section{Summary}

Figure \ref{fig:exclusion_limits} shows exclusion limits for MAGIC in the
mSUGRA plane $N_{\gamma}(E_{\gamma}>E_{\mathrm{thresh}})\langle \sigma v
\rangle$ vs. $m_{\chi}$ for the four most promising sources considered. The
nominal energy threshold $E_{thresh}$ has been assumed to be conservatively
100 GeV and has been accordingly scaled with the zenith angle observation, as
well as the telescope collection area. 

The expected fluxes are rather low and depends strongly on the innermost
density region of the DM halos considered. The detection of a DM $\gamma$-ray
signal from the Galactic Center is possible (or achievable) in case of a very
high density DM halo, like the one predicted by adiabatic contraction
processes (expected improvements on the $E_{thresh}$ allows to test a
significant portion of the SUSY parameter space). The flux measured by the
HESS (see L. Rolland in these proceedings) experiment is far above the theoretical
expectations (it extends to the TeV regime), indicating that the observed gamma
radiation is most probably not due to the annihilation of SUSY-neutralino Dark
Matter particles. Nevertheless, other models like Kaluza-Klein Dark
Matter are not ruled out. It is interesting to investigate and characterize
the observed gamma radiation to constrain the nature of the emission. In
addition, it is not excluded that a part of the flux is due to Dark Matter
annihilation. Due to the large zenith angle for Galactic Center observations,
MAGIC will have a large energy threshold but also a large collection area and
good statistics at the highest energies. Galactic Center observations are
foreseen from April to August 2005~\cite{Flix05}.

In the long term we consider Draco as a plausible candidate for Dark Matter
inspired observations. Conservative scenarios give low fluxes which are not
detectable by MAGIC in a reasonable observation time. However, there are
several factors that might enhance the expected flux from neutralino
annihilations in Draco. Other Dark Matter particles, like Kaluza-Klein
particles, may produce gamma-rays, maybe with a higher flux than those
quoted here. Draco is the most DM dominated dwarf (M/L up to 300) and an
object where no other $\gamma$-ray emission is expected. Low zenith
angle observations will preserve the nominal (low) $E_{thresh}$ of the
MAGIC telescope. Moreover, no known VHE emitters in the FOV provides no other
$\gamma$-ray sources in competition to the one predicted in the exposed DM
scenario.


\subsection{Acknowledgments}
I am grateful to the Conference Organizers for a very enjoyable week and
Conference in La Thuille and to all the members of the MAGIC Dark Matter
working group for fruitful discussions.


\section*{References}

\begin{thebibliography}{99}

\bibitem{jung96} G. Jungman, M. Kamionkowski and K. Griest, Physics Reports, 267, 195-373 (1996)

\bibitem{gas05} J. Gascon, astro-ph/0504241

\bibitem{cor05} J. Cortina for the MAGIC Collaboration, Astrophys.Space Sci. 297, 245-255 (2005)

\bibitem{nfw97} J. Navarro, C. Frenk and S. White, ApJ 490, 493 (1997)

\bibitem{Prada04} F. Prada, A. Klypin, J. Flix et al., Phys. Rev. Lett. 93, 241301 (2004)

\bibitem{gnedin04} O. Gnedin et al., Astrophys.J. 616 16-26 (2004)

\bibitem{fornengo04} N. Fornengo et al., Phys. Rev. 70 103529 (2004)

\bibitem{Lokas05} E. Lokas et al., submitted to MNRAs (2005)

\bibitem{evans04} N. W. Evans et al., Phys.Rev. D69 123501 (2004)

\bibitem{mclaugh99} D. E. McLaughlin, ApJ 512 L9 (1999)

\bibitem{Flix05} MAGIC Dark Matter Working Group, J. Flix, {\em Phd. Thesis} (2005)

\end{thebibliography}
\end{document}